\documentclass{ws-p8-50x6-00}

\begin{document}

\title{Neutron-Capture Element Trends in the Halo}

\author{C. Sneden}

\address{Department of Astronomy, University of Texas,
Austin, TX 78712, USA\\E-mail chris@verdi.as.utexas.edu}

\author{J. J. Cowan}

\address{Department of Physics and Astronomy, University of Oklahoma,
Norman, OK 73019, USA\\E-mail: cowan@physast.nhn.ou.edu}

\author{J. W. Truran}

\address{Department of Astronomy \& Astrophysics, University of Chicago, 
Chicago, IL 60637, USA\\E-mail: truran@nova.uchicago.edu}

\maketitle

\abstracts{
In a brief review of abundances neutron-capture elements (Z~$>$~30)
in metal-poor halo stars, attention is called to their star-to-star scatter, 
the dominance of $r$-process synthesis at lowest metallicities, 
the puzzle of the lighter members of this element group, and the 
possibility of a better $s$-/$s$-process discriminant.
}

\section{Introduction}

Most isotopes of elements with atomic numbers $\rm Z>30$ are
synthesized via neutron capture reactions.  
These ``n-capture" elements are the majority of the periodic table.  
In the so-called $s$-process, neutron fluxes are small enough to allow 
$\beta$-decays to occur between successive neutron captures, and 
element buildup proceeds along the valley of $\beta$-stability.  
In the $r$-process, huge but short-lived neutron fluxes overwhelm 
$\beta$-decays, creating very neutron-rich isotopes out to 
the neutron drip line.
Then multiple $\beta$-decays drive the nuclei back to the valley 
of $\beta$-stability.  
The final isotopic mixes will be very different in $r$- and $s$-process 
synthesis episodes, as will the elemental abundances summed over the isotopes.  
The $s$-process mainly occurs in the helium shell burning phases of 
low-intermediate mass AGB stars, while the $r$-process is probably associated 
with the explosive deaths of high mass stars.  
Thus, $r$- and $s$-process elemental abundance variations with metallicity 
should trace the contributions of different mass ranges of stars 
over the Galaxy's history.

Detailed comparisons of solar system meteoritic abundances of n-capture 
isotopes (Cameron\cite{Ca82}, K$\ddot{a}$ppeler {\it et al.}\cite{Ka89}) 
have yielded 
accurate breakdowns into $r$- and $s$-process parts of each isotope.  
But stellar spectroscopy generally cannot resolve the finely split isotopic
absorption components of atomic transitions, so elemental $r$- and
$s$-process abundances summed over the isotopes have been computed 
for solar system material by Burris {\it et al.}\cite{Bu00},
and references therein.  
No $n$-capture element with Z~$\leq$~83 can be identified solely
with the $r$- or $s$-process but some have been clearly dominated by a 
single synthesis mode.  
For example, Ba and Ce have $s$-process fractions, and Eu, Gd, and Dy have
$r$-process fractions greater than 80\% in solar system material.
Such elements are usually labeled as $r$-process or $s$-process,
regardless of their synthesis history in non-solar Galactic material.
 
Here we will comment on several aspects of observed 
$n$-capture abundance distributions of metal-poor stars: 
(a) the bulk $n$-capture abundance levels; (b) the relative $s$-/$r$-process 
dominance among heavier $n$-capture elements in the lowest metallicity 
stars, (c) the best spectroscopic indicator of those ratios, and (d) the
difficulty in ascribing a nucleosynthetic origin for the lighter 
$n$-capture elements.  
Theoretical interpretation of these observational results will be 
considered in a companion paper by Cowan {\it et al.} in this volume.

\section{Variations in Bulk $n$-Capture Abundance Levels}

Early large-sample surveys of $n$-capture elements in metal-poor stars 
({\it e.g.} Luck and Bond\cite{LB85}, Gilroy {\it et al.}\cite{Gi88}) 
discovered apparently significant star-to-star scatter in the ratios 
[$n$-capture/Fe], where the $n$-capture elements usually considered were 
Sr, Y, Zr, Ba, La, Nd, and Eu.  
But the observational data were usually of modest resolution (R~$<$~30,000) 
and sometimes low S/N, raising questions about the reality of the 
$n$-capture scatter.  
Recent analyses of much better spectroscopic data have decided the issue 
unambiguously in favor of $\sigma$[$n$-capture/Fe]~$>$~1 at metallicities 
[Fe/H]~$<$~--2.
The reality of these large star-to-star variations can be demonstrated
through inspection of spectra of stars with similar atmospheric parameters 
(T$_{\rm eff}$/log~$g$/[Fe/H]/v$_{\rm t}$) but enormous line 
strength differences of some $n$-capture elements.  
For example, compare the spectra of HD~115444 and HD~122563 
(Figure~1 of Westin {\it et al.}\cite{We00}; $\Delta$[Fe/H]~$\simeq$~--0.2, 
$\Delta$[Eu/H]~$\simeq$~+1.0) and those of HD~6268 and BD~+9~2870 
(Figure~3 of Burris {\it et al.}\cite{Bu00}; $\Delta$[Fe/A]~$\simeq$~0.0, 
$\Delta$[Eu/H]~$\simeq$~0.9). 
The analyses of these stars and others ({\it e.g.} McWilliam 
{\it et al.}\cite{Mc95}; Ryan {\it et al.}\cite{Ry96}) quantify these 
impressions; clearly $\sigma$[$n$-capture/Fe] grows with decreasing [Fe/H].  
To date there have been few stars with [Fe/H]~$\sim$~--3 (in which 
$n$-capture lines are weak) fully analyzed with the highest quality spectra,
and in stars with [Fe/H]$>$-2 many $n$-capture lines are saturated,
so abundances derived from them are less reliable.  
New VLT data should help at the low metallicity end of the scale; 
future $n$-capture surveys at intermediate metallicities would also help.

\section{Dominance of the $r$-Process at Lowest Metallicities}
 
Perhaps the first indication of non-solar abundance ratios of $n$-capture
elements in halo stars was the Wallerstein {\it et al.}\cite{Wa63} assertion
that the bright very metal poor ([Fe/H]~$\simeq$~--2.7) giant HD~122563 
has very small amounts of these elements compared to Fe.
Re-analysis of their data by Pagel\cite{Pa65}, and further studies of
in succeeding decades have shown that Ba is much more deficient than Eu 
in HD~122563: [Ba/Eu]~$\simeq$~--0.7.  
But attention to $s$-/$r$-process abundance mixes was really first 
brought by Spite and Spite\cite{SS78}, who found persistent deficiencies of 
Ba with respect to Eu in their sample of 11 halo stars.  
These abundances (as well as those of Y) suggested to Truran\cite{Tr81}
that ``...the observed trends follow in a natural and straightforward 
manner from the assumption that the Y and Ba in the most extreme metal-poor 
stars represent products of $r$-process nucleosynthesis."  
Since then the supporting data have improved but the basic conclusion has not.  
The observational attack has been on two fronts: (a) determination a few 
key abundance ratios in many stars, and (b) mapping the detailed 
abundance pattern in a few $n$-capture-rich stars.
 
Metal-poor stars with [$n$-capture/Fe]~$\gg$~0 have ideal spectra for
studying the entire range of $n$-capture elements, because the ubiquitous 
Fe-peak element transitions weaken substantially relative to those of
$n$-capture elements, allowing rarely-detected elements ({\it e.g.} Tb, Ho,
Hf) to be detected.  
Two prominent examples are CS~22892-052 (Sneden {\it et al.}\cite{Sn00}) and 
HD~115444 (Westin {\it et al.}\cite{We00}).  
These stars have $n$-capture abundance patterns for elements with 
Z~$\geq$~56 that are near-perfect matches to a scaled solar $r$-process 
abundance distribution.  
The presence of $s$-process synthesis cannot be detected.  
These stars' abundances suggest that regardless of the site(s) 
that are responsible for the $r$-process nuclei, they release their 
products into the ISM in a remarkably uniform pattern.\footnote{
$S$-process-rich stars do exist at very low metallicities, such as the 
``CH stars" discussed by Norris {\it et al.}\cite{No97} and McWilliam 
{\it et al.}\cite{Mc95}, but these appear to be in the minority.}
 
\begin{figure}[t]
\epsfxsize=20pc 
\hspace*{0.5in}
\epsfbox{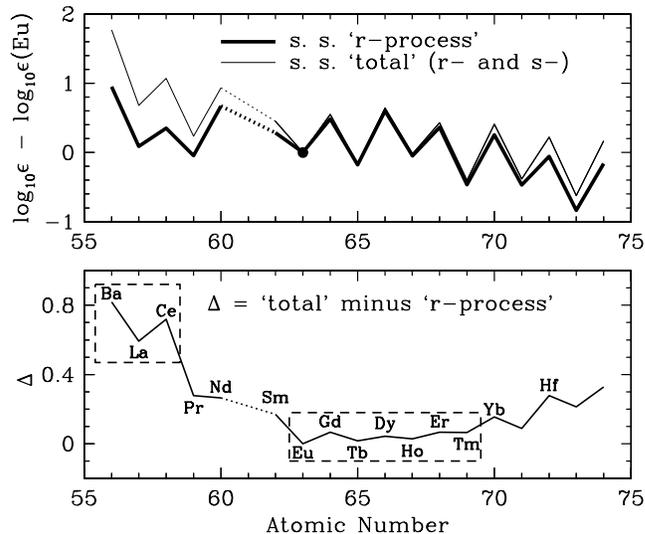} 
\caption{Top panel: total solar-system abundances and $r$-process
parts only of n-capture elements in the range 56~$\leq$~Z~$\leq$74, 
normalized to the Eu abundance.
Bottom panel: the subtraction of the two curves from the top panel.
Ratios of any element within the upper left dashed box to any element
within the lower middle dashed box should in principle be useful
indicators of the $s$-/$r$-process ratios in $n$-capture material.
\label{rs2}}
\end{figure}

Most observers simply estimate $s$-/$r$-process influence from a few 
abundance ratios of elements whose syntheses are dominated by one or the 
other mechanism in the solar system.  
In Figure~\ref{rs2} we show likely candidate elements, through 
comparison of their solar number densities log$_{\rm 10}\epsilon$. 
Clearly there should be a major abundance shift between elements
Ba$\leftrightarrow$Ce with respect to Eu$\leftrightarrow$Tm if the 
synthesis shifts
from the solar (combined $s$- and $r$-) mix to a pure $r$-process.  
In practice, this has come down to derivations of [Ba/Eu] ratios, since 
these two elements have the strongest transitions in most routinely 
accessible spectral regions.  
Nearly all large-sample $n$-capture abundance studies of low metallicity 
stars have correlated [Ba/Eu] with [Fe/H], and a consensus has arisen 
that $<$[Ba/Eu]$>$~$\sim$~0 for --2~$\leq$~[\rm {Fe/H}]~$\leq$~0, 
and then the mean ratio declines to $\sim$~--0.9 as [Fe/H]~$\sim$~--3.  
Little is known about $<$[Ba/Eu]$>$ at even lower metallicities because
the Eu transitions usually become extremely weak; it would be useful to 
detect this element even in a few stars with [Fe/H]~$<$~--3.
 
Burris {\it et al.}\cite{Bu00} have estimated the $s$-process component of
Ba in their sample of halo stars by first assuming that Eu is a 100\% 
$r$-process product and that the $r$-process part of Ba is fixed by the 
solar system $r$-process Ba/Eu ratio.
They then subtract the inferred Ba$_{\rm r-process}$ from Ba$_{\rm total}$.
The result (see their Figure~7) suggests a complete absence of the 
$s$-process for [Fe/H]~$\leq$~--2.8, and a stochastic rise to a full 
(solar-system) $s$-/$r$-process ratio by [Fe/H]~$\sim$~--2.

\begin{figure}[t]
\epsfxsize=20pc 
\hspace*{0.5in}
\epsfbox{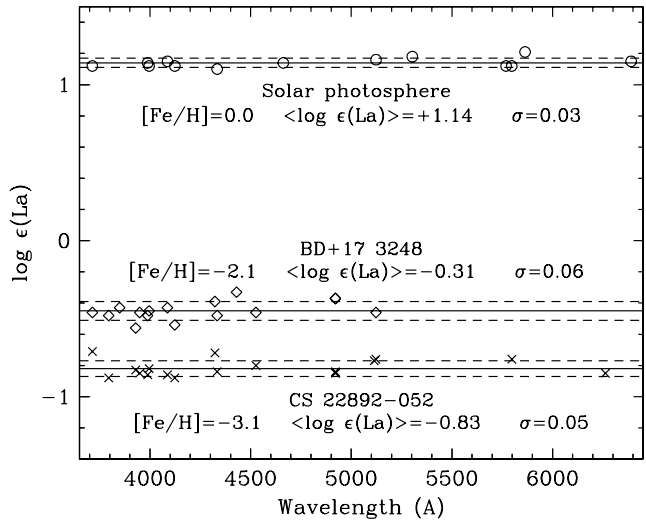} 
\caption{La abundances in the Sun and two metal-poor giants as functions
of wavelength.
The solid and dashed lines are drawn at the mean and mean$\pm\sigma$
abundances for each star.
\label{lasummary2}}
\end{figure}

Ratios of [Ba/Eu] are subject to large uncertainties, because the four
commonly employed Ba~II lines are often very strong even in metal-poor stars, 
and because these lines suffer isotopic and hyperfine structure (hfs) 
splitting, so Ba abundances are very dependent on assumed microturbulent 
velocity and isotopic abundance fractions.  
Of the alternate $s$-process abundance indicators, Ce has only weak lines, 
rendering it almost useless unless [$n$-capture/Fe]~$\gg$~0.  
But La has many weak and strong lines, and has only one stable
isotope, $^{139}\rm {La}$.  
The atomic data for La has been not the best, but Lawler 
{\it et al.}\cite{La00}have determined new accurate $gf$'s and hfs 
components for La~II.  
In Figure~\ref{lasummary2} we show results of the application of 
these lab data to the spectra of the Sun and two metal-poor but 
$n$-capture-rich stars.  
The line-to-line scatter is satisfactorily low and a reasonable mean 
abundance for all these stars is obtained.  
In preliminary tests we have derived La/Eu abundance ratios for these 
and a few other stars, finding excellent accord with the predicted pure
$r$-process ratio at the lowest [Fe/H] values, and gradual rise to solar
system values at higher metallicities.  
We believe that La/Eu ratios can be more accurately determined than 
Ba/Eu ratios, and it is possible that La/Eu may become the standard  
$s$-/$r$-process indicator in future spectroscopic investigations 
of metal-poor stars.
 
\section{The Lighter $n$-Capture Elements}
 
Brief notice must be given of $n$-capture elements in the range 
31~$\leq$~Z~$\leq$~55.  
Until recently, the elements Rb, Sr, Y, and Zr represented the observational
situation of the entire group.  
These elements are known to present a puzzle.  
Their overall abundance levels correlate poorly with heavier (Z~$\geq$~56)
$n$-capture elements (see references cited previously).  
Neither a pure solar system $r$-process, or pure $s$-process 
(or indeed any linear combination) provides a satisfactory match to their 
observed abundances (Cowan {\it et al.}\cite{Co95}).  
New near-UV spectra for some metal-poor stars are rapidly providing 
abundance data for other elements in this atomic number range such as Ag 
(Crawford {\it et al.}\cite{Cr98}), and Sneden {\it et al.}\cite{Sn00}
report first detections of five new light $n$-capture elements in 
CS~22892-052. 
But their derived abundances do not shed further insight on the matter, and 
still there is no reasonable fit to scaled solar-system abundance patterns.  
The study of these elements should be actively pursued in the future.

\section*{Acknowledgments}
We thank all of the colleagues who have collaborated with us on various
studies of $n$-capture elements in halo stars.
This research has received support from NSF grants AST-9987162 to C.S. and
AST-9986974 to J.J.C., from DOE contract B341495 to J.W.T., and from the Space
Telescope Science Institute grant GO-8342.

\end{document}